\documentclass[12pt,longbibliography,eprint]{revtex4-1}
\usepackage{epigraph}
\usepackage{graphicx}
\usepackage{epstopdf}
\usepackage{hyperref}

\begin{document}
\preprint{V.M.}
\title{Market Dynamics vs. Statistics: Limit Order Book Example.}
\author{Vladislav Gennadievich \surname{Malyshkin}} 
\email{malyshki@ton.ioffe.ru}
\affiliation{Ioffe Institute, Politekhnicheskaya 26, St Petersburg, 194021, Russia}

\author{Ray Bakhramov} 
\email{rbakhramov@forumhedge.com}
\affiliation{Forum Asset Management LLC, 733 Third Avenue,
New York, NY 10017}

\date{March, 16, 2016}

\begin{abstract}
\begin{verbatim}
$Id: DynamicsVsStatistics.tex,v 1.75 2016/03/29 23:23:26 mal Exp $
\end{verbatim}
Commonly used limit order book attributes are empirically considered
based on NASDAQ ITCH data.
It is shown that some of them have the properties
drastically different from the ones assumed in many market dynamics study.
Because of this difference 
we propose to make a transition
from ``Statistical'' type of order book study (typical for academics)
to ``Dynamical'' type of study (typical for market practitioners).
Based on market data analysis we conclude,
that most of market dynamics information is contained in
attributes with spikes (e.g. executed trades flow $I=dv/dt$),
there is no any ``stationary case'' on the market and
typical market dynamics is a
``fast excitation and then slow relaxation'' type of behavior
with a wide distribution of excitation frequencies and relaxation times.
A computer code, providing full depth order book information and
recently executed trades is available from authors\cite{polynomialcode}.
\end{abstract}

\keywords{Liquidity Deficit, Market Dynamics, Statistics, Limit Order Book}
\maketitle

\epigraph{I disagree with the world.}{Penelope Fitzgerald}

\section{\label{intro}Introduction}

Limit order book study is a hot topic of quantitative finance\cite{2008arXiv0809.0822B,toth2009studies,eisler2012price,besson2014deal,donier2015walras,2016arXiv160206968B}.
The reason is simple: while executed trades provide information about the
``past'' execution prices, order book may, possibly, provide information
about ``future'' execution prices, what attract a lot of attention.

Most of current order book study have a goal of building
a ``statistical'' type of theory. 
This typically mean 1) an introduction of some (often implicit) time--scale.
2) Calculation of some statistical characteristics, that
is claimed to help in describing market dynamics
and, in best possible case, helps in predicting future price changes\cite{cartea2014buy}.

However, when one visit a trading floor of a major bank,
or any hedge fund (at least those the authors of this paper
have seen), nobody uses ``statistical'' type of theory
at all
to make a trading decision. Market practitioners
look not on statistical parameters,
they typically talk over the phone to each other and, most of them,
watch time--dependent variables (e.g. prices,
moving averages, bond spreads, volatility charts, etc.) trying to make a trading decision.
Only after the decision is already made a
statistical theory of some kind is often deployed to estimate
a proper hedge ratio or possible risk. Same thing we have observed with
HFT trading, where an algorithms is typically represented as a
``hardwired'' trading idea
and statistical characteristics are only used for e.g. thresholds
calculation.

There is a different mindset between academical studies,
that is of ``statistical'' type,
and market practitioners mindset, that is of ``dynamical'' type.
Some market practitioners activity,
such as watching all day prices and moving averages,
may be worthless, (we believe that
price--information is insufficient for successful trading\cite{2015arXiv151005510G}).
Despite useless, or in some cases even counterproductive,
 market practitioners activity,
we, nevertheless, think that an effort should be made
to bring academic studies
close to real markets,
and the different mindset issue should be addressed.

The first simple step in this direction  is made
in this paper: we are going to plot several time--dependent
characteristics of real market data to debunk
some common in academic circles opinions about limit order book.
This may look as ``trolling'': in our previous paper\cite{2016arXiv160204423G}
we have thrown away the ``Supply and Demand'',
and now throw away the ``Statistics''. But this is the life of 
real markets. During all our
more than 40 years of combined ``hedge fund experience''
1) nobody used supplied and demand 2) statistics was used
only to justify already made decision.

\section{\label{liquiddef} Liquidity Deficit}
Before we go ahead with charts related to order book study
we provide brief introduction to our market dynamics
theory\cite{2015arXiv151005510G}, because we are going to
use it for explanations.
The execution rate,
\begin{eqnarray}
  I&=&\frac{dv}{dt}\label{I}
\end{eqnarray}
the number of shares traded per unit time,
is the key concept of our theory.
Numerical estimation of (\ref{I})
can be done by  using either some crude techique such as ``sliding window'', 
or, much better option, by an application of Radon--Nikodym
derivatives\cite{kolmogorovFA} on two measures $dv$ and $dt$,
see \cite{2015arXiv151005510G} for details of numerical estimation of
it and the calculation of 
$I_0$ ($I$ ``now'') with possible thresholds for $I_0$
(using either boundary conditions for the probability state, or a projection of
the state ``now'' to the states of minimal and maximal $I$).

As we established empirically \cite{2015arXiv151005510G,2016arXiv160204423G}
the trading rate (\ref{I}) is the most important
characteristics affecting price dynamics.
This is especially evident in a quasi-stationary case \cite{2016arXiv160204423G},
where high price changes are observed on high values of $I$, but with little volume traded.

The (\ref{I}) with thresholds define
what market practitioners call ``slow'' and ``fast'' markets.
In this sense we agree with the Ref. \cite{mandelbrot2014misbehavior} B. Mandelbrot and R. Hudson:

``In fractal analysis, time is flexible. The multifractal model describes markets as deforming time—expanding it here, contracting it there. The more dramatic the price changes, the more the trading time-scale expands. The duller the price chart, the slower runs the market clock. Some researchers have tried linking this concept to trading volume: High volume equals fast trading time. That is a connection not yet established, and it need not be. Time deformation is a mathematical convenience, handy for analyzing the market; and it also happens to fit our subjective experience. Time does not run in a straight line, like the markings on a wooden ruler. It stretches and shrinks, as if the ruler were made of balloon rubber. This is true in daily life: We perk up during high drama, nod off when bored. Markets do the same.''

But with one very important adjustment.
Link how the ``the trading time-scale expands'',
not to the ``trading volume'', but to the ``trading rate'' the $I=dv/dt$ from (\ref{I}).
In our earlier papers\cite{2015arXiv151005510G,2016arXiv160204423G}
we linked the high price volatility periods to the periods of high $I$ value,
what is especially evident in a quasi-stationary case.

In \cite{besson2014deal}
authors emphasize that
information propagates very fast and market makers update their quotes immediately.
We think that for US equity and fixed income markets the main source of this
information is the fact that
market participants ``feel'' the trading rate $I$ and once the it become
large -- they immediately start adjusting their quotes, thus making the large price movements. For these markets the trading rate $I$ can be considered as
the ``driving force''.
Market observation show, that prices react on trading rate much stronger than on, say, news or other events.
Some traders tried to estimate this trading rate ``dynamic impact'' experimentally\cite{EyakushevComm}
on NASDAQ exchange (one can move not very liquid stock a lot with just half a million in capital)
but such experiments are really costly.
For other markets (e.g. US real estate market) we do not have
an answer whether the $I$ (e.g. the total price of houses (or their number) sold in unit time)
is the driving force of the market
or is a consequence of some other, unknown factor. It may be both.

We believe the trading rate $I$ from (\ref{I}) is the key
factor defining US equity market dynamics and we will present it
in all the charts below.

\section{\label{ordcharts} Order Book Charts}

Despite wide attention to order book study,
the interpretation of attributes, used in these theories,
often looks very much as ``a falsis principiis proficisci'',
or, to quote the Ref. \cite{taleb2014precautionary}:

``Today many mathematical or conceptual models that are
claimed to be rigorous are based upon unvalidated and
incorrect assumptions. Such models are rational in the sense that
they are logically derived from their assumptions, except that it
is the modeler who is using an incomplete representation of the
reality.''

Below we provide, as a simple example, three commonly
used order book attributes and discuss their role in market dynamics.
In all the Figs. \ref{fig:pbuypsell}, \ref{fig:vbuyvsell} and \ref{fig:tbuytsell}
we present the data, obtained from NASDAQ ITCH data feed\cite{itchfeed},
for the AAPL stock on September, 20, 2012 round 10am.
The time on x axis is in decimal fraction of an hour, e.g. 9.75 mean 9:45am.
For a reference the execution flow $I_0$ is presented on each chart (Scaled to fit the chart. Calculated in Shifted Legendre basis with
$7$ elements in basis and $\tau$=128sec. See Ref. \cite{2015arXiv151005510G}, Section II.B
for numerical calculation method.).
The $I$ is believed to be the driving force of the market and for this reason
it is presented on all the charts.

\subsection{\label{midprice}Midprice Value}

\begin{figure}
\includegraphics[width=18cm]{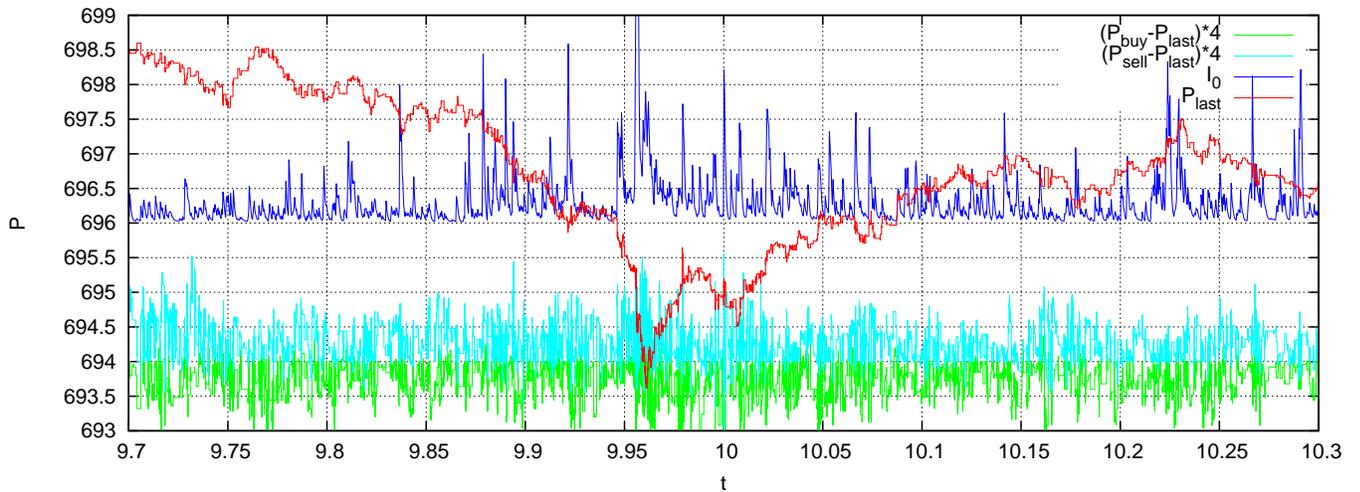}
\caption{\label{fig:pbuypsell} 
  $P_{sell}-P_{last}$ and $P_{buy}-P_{last}$ (scaled to fit the chart).
  The $P_{last}$ is last execution price and  $P_{buy}$ and $P_{sell}$ are the best order book bid \& offer.
}
\end{figure}

The midprice,
\begin{eqnarray}
  P_m&=&\frac{P_{sell}+P_{buy}}{2} \label{Pmidprice}
\end{eqnarray}
where $P_{buy}$ and $P_{sell}$ are the best order book bid \& offer,
is often considered as an important price to be used
in order book dynamics models.
In the Fig. \ref{fig:pbuypsell}
we present the $P_{sell}-P_{last}$ and $P_{buy}-P_{last}$.
These fluctuate only within about a spread value, but it is
clear that during the periods of high value of $I_0$ the trading
typically occurs on one side of the book and the spread can either
growths or shrink to minimal value when a ``maximal $I$'' price is reached
and buyers and sellers send to the exchange whatever liquidity
they have. In the last case the two types of matching
``market sell matched limit buy'' and ``market buy matched limit sell''
become almost identical.
This make the midprice (\ref{Pmidprice})
irrelevant during most interesting moments of high $I_0$
(correspond to the moments of high price volatility).
We also emphasized in \cite{2015arXiv151005510G},
that current NASDAQ  fee structure make order book
manipulation close to free. In addition to that,
if you talk to market practitioners, and our personal trading experience confirms this,
since about 2008--2010 nobody in a sane mind
would trade according to the $P_{buy}$ and $P_{sell}$ levels of the order book.
Actual exchange liquidity is typically significantly better,
than the $P_{buy}$ and $P_{sell}$ levels. If you send a ``buy'' order
slightly lower than $P_{sell}$ or ``sell'' order slightly higher than  $P_{buy}$,
then this order typically gets almost immediately matched
by a market order coming. This means that actual best buy/sell prices
are substantially better than the order book reports and to
trade market order according to order book best levels guarantee to lose
at least a few cents in price and lose the rebate
paid by the exchange for executed limit orders, compared to an alternative
of posting limit order at price few cents better than order book best price level.
Moreover, experiments show, that for high liquidity stocks, buy order with price 
$\min(P_{sell},P_{last}+\delta)$
and sell orders with price $\max(P_{buy},P_{last}-\delta)$
get almost immediately executed.
The value of $\delta$ is about few cents (depend on stock liquidity, and spread).

Another concept we have widely seen in academic publications
is a concept of ``patient'' and ``impatient'' traders.
We think that this concept has nothing to do with reality.
The goal of all traders is to get orders executed at
a good price. When some trader put a limit
order deep inside order book --- this is not because
this trader is ``patient''. It is very opposite: a trader
put an order deep inside the order book only
because this trader want be the first (very impatient) when
the price reaches that level.
We believe, that
 exchange trading since at least 2010
is not very much different from a ``dark pool'' trading:
you send an order at some price --- you may or may not get it executed,
and the order book information is of little value.
Typical exchange order execution pattern is this:
an order came at some price in between book "best buy" and "best sell",
spend almost no time in the order book,
then it either get almost immediately executed or cancelled.
The ratio observed is that more than 90\% of orders
being at best price level at some time end up being cancelled\cite{nasdaqord,2015arXiv151005510G}.
The order book contains mostly "old" orders, having no much effect on execution rate.
This make us to conclude that since at least 2010 available
from exchange order book best price levels are way too conservative,
and actual ``exchange'' info is not that much different from a ``dark pool'' info.
This mean that the order book spread and the midprice (\ref{Pmidprice}) are almost useless.

\subsection{\label{voldis}Volume Disbalance}

\begin{figure}
\includegraphics[width=18cm]{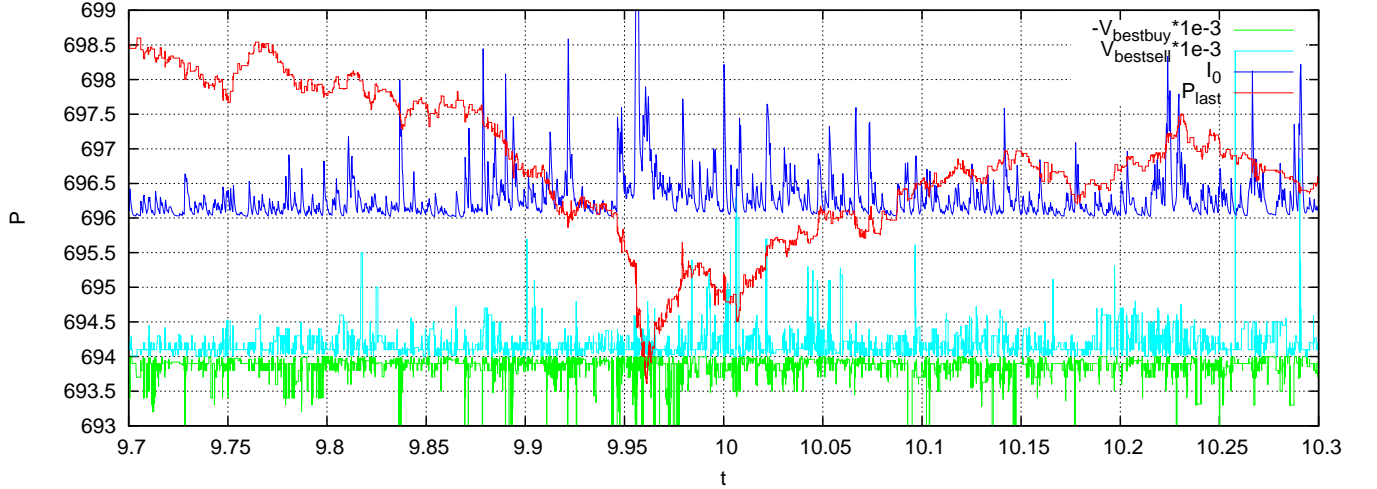}
\caption{\label{fig:vbuyvsell} 
  The values of book volume on best buy/sell price levels (scaled to fit the chart).
}
\end{figure}

Order book best levels volume relative disbalance (\ref{vsvbdisb})
is often considered to be an important attribute for market dynamics theories.
\begin{eqnarray}
  \eta_{disbalance}&=&\frac{V_{best\, sell}-V_{best\, buy}}{V_{best\, sell}+V_{best\, buy}}
  \label{vsvbdisb}
\end{eqnarray}
This attribute is commonly  interpreted as defining the
asymmetry of ``price impact'': 
the side with the biggest best level volume providing ``price support'';
price move in this direction requires a bigger volume traded.
In the Fig. \ref{fig:vbuyvsell} we present the absolute
value of order book best price level volume (scaled to fit the chart).
First, what is very clear from the charts, is
the spikes in volume on best price levels,
that correlate
with high value of trading rate $I$ (\ref{I}).
Another very clear observation from this chart
is that large best price level volume can have a different effect
on price dynamics. Check, for instance, time interval
between $9.95$ and $10.0$. There large best bid volume
can serve both as ``price support'' and as ``liquidity attractor''.
For ``liquidity attractor'' market participants see
large liquidity on the best price level,
and, if this price is acceptable to them,
take it immediately, considering best level liquidity
as an opportunity. This ``liquidity attractor'' effect is opposite
to the ``price support''. Ideologically the situation
is similar to the situation with large price change.
What can happen after asset price strong move in some direction.
It can be both: bounce back or follow the trend.
Very similar situations is here. Large order book best price level volume
can cause both: ``price support''(bounce back)
and ``liquidity attractor''(trend following),
not to mention the most common outcome --- orders cancellation.

We think that widely used in academic study
best price level volume disbalance indicator (\ref{vsvbdisb})
has a bad normalizing (spikes are the most informative features)
and volume disbalance should be interpreted ambivalently:
both as ``price support'' and as ``liquidity attractor''.

\subsection{\label{taudis}Time In Order Book}

The best price level volume value is not a very good
indicator for the reason market participants typically keep
the order book volume at some moderate level and update
it as long as execution goes
for the reason not to reveal their actual intentions.
In this sense average time (calculated as the difference between 
current time and limit order origination time, weighted with order size)
on best price level
can serve as a better than volume indicator of liquidity influx.

\begin{figure}
\includegraphics[width=18cm]{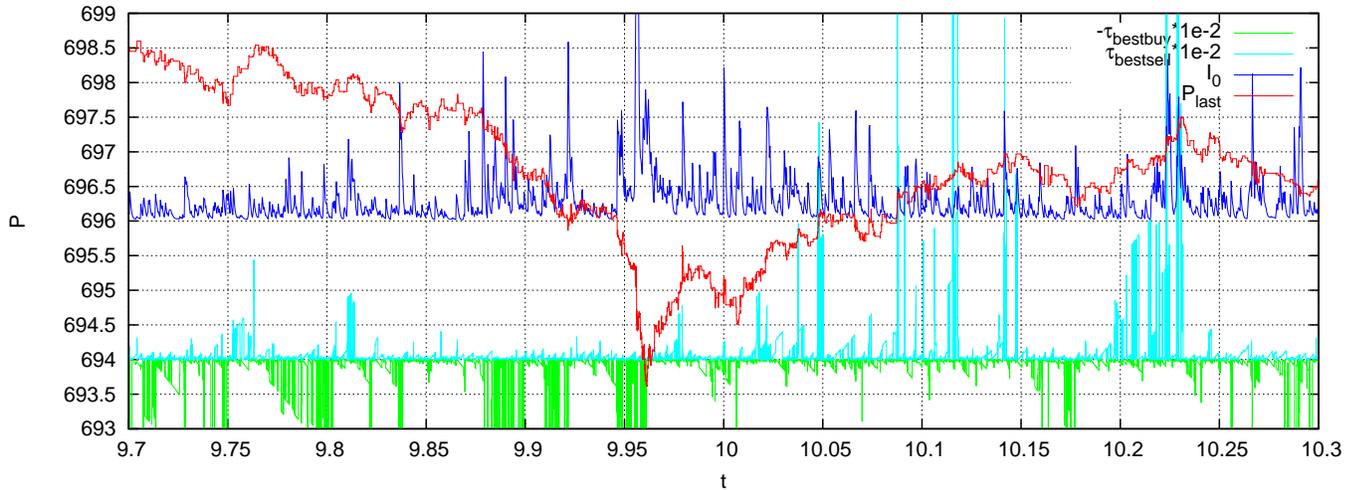}
\caption{\label{fig:tbuytsell} 
  Time spend in order book for the orders on best buy/sell price level.
}
\end{figure}

It should be noted, that NASDAQ exchange also has non--displayable
(previously called ``hidden'')
orders, that cost more to place and are not reported in order book.
They are only
reported in execution. Executed non--displayable order id was
actually available before October 6, 2010,
what allowed to interpolate hidden order origination time,
but after this date NASDAQ
broadcast 0 as hidden order id.
In addition to that
effective July 14, 2014, NASDAQ
does not report matching type (market--buy matched limit--sell
or market--sell matched limit--buy)
for non--displayable orders.
See Appendix A of Ref. \cite{itchfeed}.
For these reasons
all the order data analyzed in this paper actually misses
all hidden orders information.

In Fig. \ref{fig:tbuytsell} for the orders on the best buy/sell level
the average time spent in order book is presented.
This time has spikes (similar to volume spikes, but more clear ones),
and these spikes correlate with large $I$ (\ref{I}).
The picture is similar to the volume disbalance of subsection (\ref{voldis}),
but the result is less ambivalent between ``price support'' and ``liquidity attractor''.
This can be interpreted in a way similar to ``trading volume vs trading rate''
of Ref. \cite{2016arXiv160204423G}, here it would be the 
``volume at best price level vs orders influx (proportional to inverse time)''.
As we emphasized in Ref. \cite{2015arXiv151005510G}
the time spent in order book is one of a few good indicators  obtained
out of the order book data.
The reasons may be: difficulty to calculate (the order book is required),
harder to manipulate (time is not possible to ``fix'', but the volume is easy)
and being rate--like attribute.

\section{\label{disc}Discussion}
In this paper we emphasize the importance
for academic community to make a transition from
``Statistical'' to ``Dynamical'' type of order book study
to bring order book study more in line with market practitioners activity.

We propose the following steps in this direction :
\begin{itemize}
\item Stop making distribution charts and start
  making time--dependent charts. As a contribution
  to this pathway we provide computer code, see Appendix \ref{code}
  as a reference implementation.
  Take e.g. com/ polytechnik/ algorithms/ ExampleBookDisbalance.java
  and modify the ``processData'' method, that is called on every book modification event,
  and has three arrays as the arguments: recently executed trades,
  order book buy side orders, order book sell side orders.
  It is seductively easy to make some statistics out of these arrays,
  but try not to do this, and, instead, do plain output of the attributes calculated.
  Then look at them as time--dependent values.

\item Stop normalizing calculated values. Do not divide on standard deviation
  or other ``ad hoc'' selected scale.
  For example, used in numerous study the sell--buy book order best level volume
  relative disbalance (\ref{vsvbdisb}) we think can be used only by those who have
  never seen actual order book volume at best price level
  (or those who have no risk to lose real money if the market goes against your positions).
  As we emphasized in the subsection \ref{voldis} actual volume
  at best price level have huge spikes, the changes are often of several
  orders of magnitude. The most important information is contained  in
  these spikes, because they determine the liquidity available.
  Dividing the volume by the absolute value (\ref{vsvbdisb})
  remove the spikes and make the attribute worthless.
  Moreover, we think that most market study
  should be directed toward the ``attributes with spikes'' as they define the dynamics,
  e.g. the trading rate (\ref{I}), or time the order spent in the order book,
  considered in subsection \ref{taudis}.
  The ``attributes with spikes'' are fundamental to
  e.g. Limit Order Book structure or execution flow:
  there is no any stationary state for these variables;
  the data supported fact is that most of observable variables are
  severely non--stationary and represent a combination of
  ``fast excitation and then slow relaxation'' type of behavior
  with a wide distribution
  of excitation frequencies and relaxation times.  
  Once the ``attributes with spikes'' are identified --- the threshold
  for their estimation of ``low'' and ``high'' value is required.
  This is a complex problem, no simple thing with fixed time scale statistics
  (e.g. standard deviation)
  would ever work. In our \cite{2015arXiv151005510G} paper we proposed
  a ``probability states'' --type of answer,
  an application of that was demonstrated there for the calculation of $I$ thresholds.
\item
  Stop linking model quality with the quality of price prediction.
  Predicted price typically provide completely meaningless result (but often very good Sharpe Ratio).
  Because of price prediction errors the trading strategy can give both positive and negative P\&L.
  We think that model quality should be based on the quality of P\&L prediction.
  P\&L dynamics include not only price dynamics, but also trader actions.
  There is ``a posse ad esse'' question  about the possibility
  of future price prediction. We believe that price prediction
  is possible only in some seldom moments.
  The key element of any P\&L trading strategy  is an existence
  of at least four signals (trader actions): ``Enter Long'', ``Exit Long (sell existing long)'', ``Enter Short'', ``Exit Short (buy to cover)''. 
  A profitable P\&L trading strategy
  should open the position during the time, when future market movement
  can be predicted (enter condition) and closing the position when future direction is uncertain (exit condition).
  In our work \cite{2015arXiv151005510G} we linked opening position (but the decision about ``long'' or ``short''
  is still problematic) with liquidity deficit event and closing the position
  with liquidity excess\footnote{
    Note, that commonly used trading strategies,
    for which enter/exit conditions are different only in threshold values
    can never work as a  P\&L trading strategy.
    This is a trivial illustrative example of a ``mean reverse strategy'':
    given price $P$, moving average $P_A$,
    and enter and exit thresholds, satisfying $\sigma_{enter}>\sigma_{exit}$.
    Then typical conditions are:\\
    ``Enter Long'':$P-P_A<-\sigma_{enter}$;
    ``Exit Long'':$P-P_A>\sigma_{exit}$; \\
    ``Enter Short'':$P-P_A>\sigma_{enter}$;
    ``Exit Short'':$P-P_A<-\sigma_{exit}$. \\
    For a P\&L trading strategy
    enter condition must
    identify the moments when the price can be predicted,
    and exit condition must identify the moments,
    when the price cannot be predicted and the position
    must be closed to avoid price uncertainty risk
    (otherwise a price move against position held,
    that lead to a catastrophic P\&L loss
    is guaranteed to happen sooner or later).
    The problem with these, just threshold--different enter/exit conditions,
    such as e.g.
    ``Enter Long'':$P-P_A<-\sigma_{enter}$ and
    ``Exit Short'':$P-P_A<-\sigma_{exit}$
    --- they are essentially the same.
    They do not carry any information,
    that qualitatively distinguish possibility or impossibility 
    of price prediction.
    In  \cite{2015arXiv151005510G} we used
    low/high value of the execution rate $I$ as the criteria
    distinguishing enter/exit condition.
    This $I$--based criteria can determine possibility or impossibility 
    of price prediction.
    It is qualitatively different from the direction determining criteria.
    A P\&L trading strategy is ideologically similar (but not identical,
    because P\&L trading do have a directional component)
    to volatility trading strategy.
    Take a position during low volatility and close it during high volatility
    to avoid unexpected price move.
    Because price high volatility correspond to high $I$ (the $I$ is the driving force of the market),
    the position held must be closed when the value of $I$ is high. Consequently,
    it should be opened when the  $I$ is low.
    
    The $I$, along with limit order cancellation flow,
    describe what the authors of \cite{besson2014deal}
    call the ``liquidity consumption''. The balance of
    ``liquidity offer'' (orders influx to the Limit Order Book)
    and ``liquidity consumption'' is currently an active research subject.
    Many authors made attempts to obtain large time scale price dynamic
    out of the disbalance of ``liquidity offer'' and ``liquidity consumption''.
    Our numerical experiments (the code is very similar to
    the one described in Appendix \ref{code}),
    show that ``liquidity offer'' and ``liquidity consumption'' flows
    have almost identical timing (check the position of the spikes in flows plotted as
    time--dependent charts, do not use any statistics).
    This gives possible prediction time, obtained from flows--disbalance,
    not exceeding milliseconds scale (the ITCH data from \cite{itchfeed}
    provide time resolution of one nanosecond).
    At such a small time scale there is very little liquidity
    available and a fierce competition for it among the HFT firms,
    make such a small timescales unworkable for 
    a ``regular'' HFT firm.
    A larger time scale is required. The approach of subsection \ref{taudis},
    where the time spent in the order book is demonstrated as a possible
    approach to reach a ``workable'' time scale.
    Another option to obtain a ``workable'' time scale
    is to use the dynamic equation $I\to\max$, that we introduced in Ref. \cite{2015arXiv151005510G}.
    Knowing the probability state $\psi(t)$, maximizing the
    functional $<I\psi^2(t)>/<\psi^2(t)>$ (stable estimator\cite{malha,malyshkin2015norm}),
    is equivalent to knowing the ``proper'' time scale.
    Anyway, for any attainable to a ``regular'' HFT firm time scale, the
    ``liquidity offer'' and the ``liquidity consumption''
    can often be considered as redundant,
    and the execution rate $I$
    is sufficient to be considered as a proxy for both.
    The situation here is similar to our alternative to Supply--Demand theory,
    the Liquidity Deficit theory\cite{2016arXiv160204423G},
    where supply and demand are considered as always matched,
    but the rate of their matching $I$  varies.
    The \cite{2016arXiv160204423G} demonstrate
    this on a large time scale (one trading day).
    Numerical experiment with the code
    from Appendix \ref{code} and NASDAQ ITCH data\cite{itchfeed}
    can demonstrate ``liquidity offer'' and ``liquidity consumption''
    matching down to milliseconds scale (in a sense time difference
    in spikes position).
  }.
Other approaches to opening and closing positions
  can be used, but what is important for any trading strategy is the
  separation of price movement and trader actions.
  Together they give the P\&L,
  that is ``the ultimate criteria'' of any trading strategy quality.
  In this sense the P\&L dynamics is not ``ad captandum vulgus'',
  what some academics was telling us. And not only academics,
  the backtesting process in some practical automated trading machines
  we have seen was actually an estimation of price prediction quality.
  We believe that the P\&L dynamics, that separates trader actions
  and price movement, should
  be the fundamental topic of any market dynamics study.
\end{itemize}

\begin{acknowledgments}  
  Vladislav Malyshkin would like to thank
  Alberto Bicci and Charles--Albert Lehalle 
  for fruitful discussions.
\end{acknowledgments}

\appendix
\section{\label{code}Code implementation}
Computer code implementing the algorithms is available\cite{polynomialcode}.
The code is java written. To reproduce the results
follow these steps:
\begin{itemize}
\item Install java 1.8 or later.
\item Download from \cite{polynomialcode}  the data file S092012-v41.txt.gz
  and code archive SupplyDemandQuasiStationary.zip.
\item Decompress the code and recompile it.
\begin{verbatim}
    unzip SupplyDemandQuasiStationary.zip
    javac -g com/polytechnik/*/*java
\end{verbatim}
\item
  Run the command:
\begin{verbatim}
java com/polytechnik/algorithms/ExampleBookDisbalance \
       S092012-v41.txt.gz AAPL >book_aapl.csv
\end{verbatim}
to extract order book edges information on every order book modification event.
The output include last executed and book best prices, book best level volume,
and book best level time. Book edges approximation (e.g. volume via Christoffel function) is also presented (for approximation  the book orders
are cut at high enough price level (about \$1 from the best price),
then 10--points Gauss--Radau quadrature is built on this price--volume
distribution and the weight (equal to Christoffel function value)
at best price give best price level volume interpolation).
If one does not need Christoffel function (for the volume) and Radon--Nikodym
(for $\tau$) book edge approximation, then
the class com/ polytechnik/ itch/ DumpData2Trader.java
can be used instead of the
com/ polytechnik/ algorithms/ ExampleBookDisbalance.java.
\item
The ``processData'' method, that is called on every book modification event,
 has three arrays as the arguments: recently executed trades,
 order book buy side orders, order book sell side orders; the method
 can be modified to produce other output.
  
\end{itemize}

\bibliography{LD}

\end{document}